# The first law of thermodynamics in hydrodynamic steady and unsteady flows: The internal energy as a function of parameters of state

Konrad Giżyński[1], Karol Makuch[1], Jan Paczesny[1], Paweł Żuk[1,2], Anna Maciołek[1,3]

and Robert Hołyst[1*]

[1]Institute of Physical Chemistry, Polish Academy of Sciences
Kasprzaka 44/52, 01-224 Warsaw, Poland

[2]Department of Physics, Lancaster University
Lancaster LA1 4YB, UK

[3]Max-Planck-Institut für Intelligente Systeme
Stuttgart, Heisenbergstr. 3, D-70569 Stuttgart, Germany

**ABSTRACT**

We studied planar compressible flows of ideal gas as models of a non-equilibrium thermodynamic system. We demonstrate that internal energy $U(S^*, V, N)$ of such systems in stationary and non-stationary states is the function of only three parameters of state, i.e. non-equilibrium entropy $S^*$, volume $V$ and number of particles $N$ in the system. Upon transition between different states, the system obeys the first thermodynamic law, i.e. $dU = T^* dS^* - p^* dV + \mu^* dN$, where $U = \frac{3}{2} NRT^*$ and $p^*V = NRT^*$. Placing a cylinder inside the channel, we find that $U$ depends on the location of the cylinder $y_c$ only *via* the parameters of state, i.e. $U(S^*(y_c), V, N(y_c))$ at $V = const$. Moreover, when the flow around the cylinder becomes unstable, and velocity, pressure, and density start to oscillate as a function of time, *t*, $U$ depends on *t* only *via* the parameters of state, i.e. $U(S^*(t), V, N(t))$ for $V = const$. These examples show that such a form of internal energy is robust and does not depend on the particular boundary conditions even in the unsteady flow.

**I. INTRODUCTION**

Non-equilibrium thermodynamics (NET) addresses systems manifesting significant macroscopic energy, mass, and momentum fluxes indicative of their departure from thermodynamic equilibrium [1]. Such dynamic behaviours defy conventional descriptions offered by classical thermodynamics. NET furnishes a conceptual framework for comprehending these intricate systems, elucidating their energy transfer pathways and interactions with the surrounding environment. The practical significance of NET is underscored by its applicability to a diverse array of real-life systems, spanning physical, chemical, and biological domains, including fluid dynamics, chemical reactions, atmospheric processes, and living organisms. Several theories have been advanced to characterise the behaviour of NET systems. Noteworthy among these are Extended Irreversible Thermodynamics [2], which introduces higher-order terms into the entropy production equation, linear non-equilibrium thermodynamics [3] applicable close to equilibrium scenarios, and Prigogine's Theory of Dissipative Structures [4], which focuses on the spontaneous emergence of structures during energy and matter exchange. NET concepts have been also successfully applied to systems with a shear-flow [5–12]. Our group has developed a theory concentrating on the global internal energy accumulated in NET systems in steady states [13–22]. We have demonstrated through a series of publications that classical thermodynamic formalism can be extended to NET systems by mapping distributed properties, such as density or temperature, to averaged, effective values. This approach, both mathematically straightforward and leveraging the well-established methods of thermodynamic equilibrium, holds promise for predicting state transitions in complex NET systems.

Our previous studies focused on closed systems where energy is exchanged with the environment only in the form of heat and work. We showed analytically and rigorously [23] that for the ideal gas in the stationary state subjected to the heat flow, the internal energy of the ideal gas $U(S^*, V, N)$ is the function of only three global parameters of state: the non-equilibrium entropy, $S^*$, volume, $V$, and the number of particles $N$. We identified $T^* dS^*$ with a net heat that flow into the system that changes the internal energy. The change of the internal energy in any process moving the system from one stationary state to another, infinitesimally close, stationary state in the absence of any external field is given by $dU = T^* dS^* - p^* dV + \mu^* dN$.

In this paper, we extend the concept of NET parameters of state to open systems that exchange matter with environment, precisely the planar flows of the ideal gas. This flow type is well-studied. At the same time, its thermodynamic attributes are well defined through the equations of state. In our previous study on Poiseuille flow, we demonstrated that its internal energy is a function of four parameters that combine the control parameters: $T_0$, $\Delta p$ and $p_0$ and the size of the system [16]. Here, we examine, in analogy to our previous study of the gas in the heat flow, the internal energy as a function of parameters of state $U(S^*, V, N)$. The thermodynamic form would not depend on the number of boundary conditions or changes to the flowing gas. We subject the system to various changes in external parameters and investigate the relation $dU = T^* dS^* - p^* dV + \mu^* dN$ during these changes. Additionally, we placed a cylinder in the planar channel and checked how the internal energy depends on the location of this cylinder. Finally and most importantly, we demonstrate that the same methodology can also be applied to unsteady flows where all parameters in the system oscillate in time, $t$. Here, the boundary parameters are constant. All these checks will tell us how robust the proposed form of internal energy is as a function of only three parameters of state. The paper is organised as follows: Section II describes the hydrodynamic model based on the Navier-Stokes equations and the equations of state. Section III, containing results, is divided into three subsections corresponding to the three considered flows. We conclude our observations in Section IV.

## II. MODEL AND METHODOLOGY

We study the planar compressible flow of a monoatomic ideal gas through a rectangular channel of length $L_x$ and witdth $L_y$. The system is translationally invariant in *z*-direction. We analyse three scenarios: (i) stationary Poiseuille flow, (ii) stationary flow with a cylindrical obstacle moved in *y* direction and (iii) unsteady oscillatory flow. The three conservation laws for the mass, the momentum, and the energy governing the system have the following form [24,25]:

$$\frac{\partial \rho}{\partial t} + \nabla \cdot (\rho \vec{\mathbf{u}}) = 0, \qquad (1)$$

$$\rho \frac{\partial \vec{\mathbf{u}}}{\partial t} + \rho \vec{\mathbf{u}} \cdot \nabla \vec{\mathbf{u}} = -\nabla p + \nabla \cdot \left[ \mu \left( \nabla \vec{\mathbf{u}} + (\nabla \vec{\mathbf{u}})^T - \frac{2}{3} (\nabla \cdot \vec{\mathbf{u}}) \mathbf{I} \right) \right], \qquad (2)$$

$$\rho c_p \frac{\partial T}{\partial t} + \rho c_p \vec{\mathbf{u}} \cdot \nabla T = \nabla \cdot (k \nabla T) + \vec{\mathbf{u}} \cdot \nabla p + \left[ -\frac{2}{3} \mu \nabla \cdot \vec{\mathbf{u}} \mathbf{I} + \mu (\nabla \vec{\mathbf{u}} + (\nabla \vec{\mathbf{u}})^T) \right] : \nabla \vec{\mathbf{u}}. \qquad (3)$$

where $\rho$ is the density, $\vec{\mathbf{u}}$ is the velocity field vector, $p$ is the pressure, **I** is the identity 3-dimensional matrix, $c_p$ is the specific heat capacity at constant pressure, and $T$ is the temperature. For cases (i) and (ii), we are only interested in the stationary states so we omit time derivates in Equations 1-3.

The other two complementary equations we used are the equations of state for an ideal gas:

$$RT\rho = Mp, \tag{4}$$

$$U = c_v nMT, \tag{5}$$

where $R$ is the gas constant, $M$ is the mean molar mass, $c_v$ corresponds to specific heat capacity at constant volume, and $n$ is the number of moles of the gas. We make the assumption of the local equilibrium.

We selected Helium ($^4$He) as the working medium to represent the ideal gas with the following values of heat capacity at constant volume and pressure and mean molar mass: $c_v = \frac{3}{2}\frac{R}{M}$, $c_p = \frac{5}{2}\frac{R}{M}$, $M = 4$ g/mol. In case (i) and (ii) thermal conductivity $k$ and viscosity $\mu$ were temperature-dependent parameters. In case (i) they were described with the default COMSOL polynomial functions defined for Helium:

$$k(T) = [0.0294900023 + 5.07655059 \times 10^{-4}\,T - 4.22501605 \times 10^{-7}\,T^2 + 2.1209438 \times 10^{-10}\,T^3]\,\text{W/(mK)}, \tag{6}$$

$$\mu(T) = [3.7251756 \times 10^{-6} + 6.83450863 \times 10^{-8}\,T - 5.07299333 \times 10^{-11}\,T^2 + 2.36688744 \times 10^{-14}\,T^3]\,\text{Pa s}. \tag{7}$$

Polynomial functions describing $k$ and $\mu$ in case (ii) are given in ref. [16]. In case (iii) these were constant and equal to 1 W/(mK) and 1 × 10$^{-5}$ Pa s respectively.

Equations (1-8) were solved using COMSOL Multiphysics 6.1 with CFD [26] and Heat Transfer modules [27]. In all studied cases, we applied the same boundary conditions (BC) on the channel walls. At the inlet, we applied the fully developed flow condition prescribing tangential flow component at the inlet boundary to zero (for details see p. 183-184 from ref. [26]). For cases (i) and (ii), we fixed average pressure at $\Delta p + p_0$, where $p_0$ is the outlet pressure. In case (iii), we defined the velocity profile using parabolic equation: $\vec{\mathbf{u}}(y) = u_m y(L_y - y)/L_y^2 f_s(t)$ where $u_m$ is the mean velocity, and $f_s(t)$ is a time-dependent step function changing from 0 to 1 during the initial time of the simulation (0, 0.1 s). $f_s(t)$ enabled smooth initialisation of the flow. The inlet temperature was uniform and equal to $T_0$ in all the cases. For the sidewall, a no-slip BC ($\vec{\mathbf{u}} = 0$) was imposed, and its temperature was also uniform and kept at $T_0$. At the outlet, the pressure was static and equal to $p_0 = 1$ atm. The heat flux through the outlet is $\vec{a} \cdot \nabla T = 0$, where $\vec{a}$ is the normal vector. Additionally, in cases (ii) and (iii), the walls of the cylinders were thermally insulated with no-slip BC. A schematic representation of the system with indicated BCs is shown in Figure 1a-c, for study cases (i) to (iii), respectively. Note that in case (i), due to the symmetry of the flow, we simulated only half of the channel with a symmetry axis at y = 0. The length of the channel was fixed at $L_x = 0.5$ m. In study case (ii), we placed a cylinder with a diameter equal to $L_y/10$ at $x_c = L_x/2$. The position of the cylinder in y-direction ($y_c$) was varied. The size of the channel was fixed with $L_x = 0.5$ m and $L_y = 0.05$ m. In scenario (iii), we used the geometry from a benchmark study defined by Schaeffer and Turek [28], i.e. a cylinder with a diameter of 0.1 m fixed at $x_c = 0.2$ m, $y_c = 0.2$ m. The dimensions of the channel in this case were $L_x = 2.2$ m and $L_y = 0.41$ m.

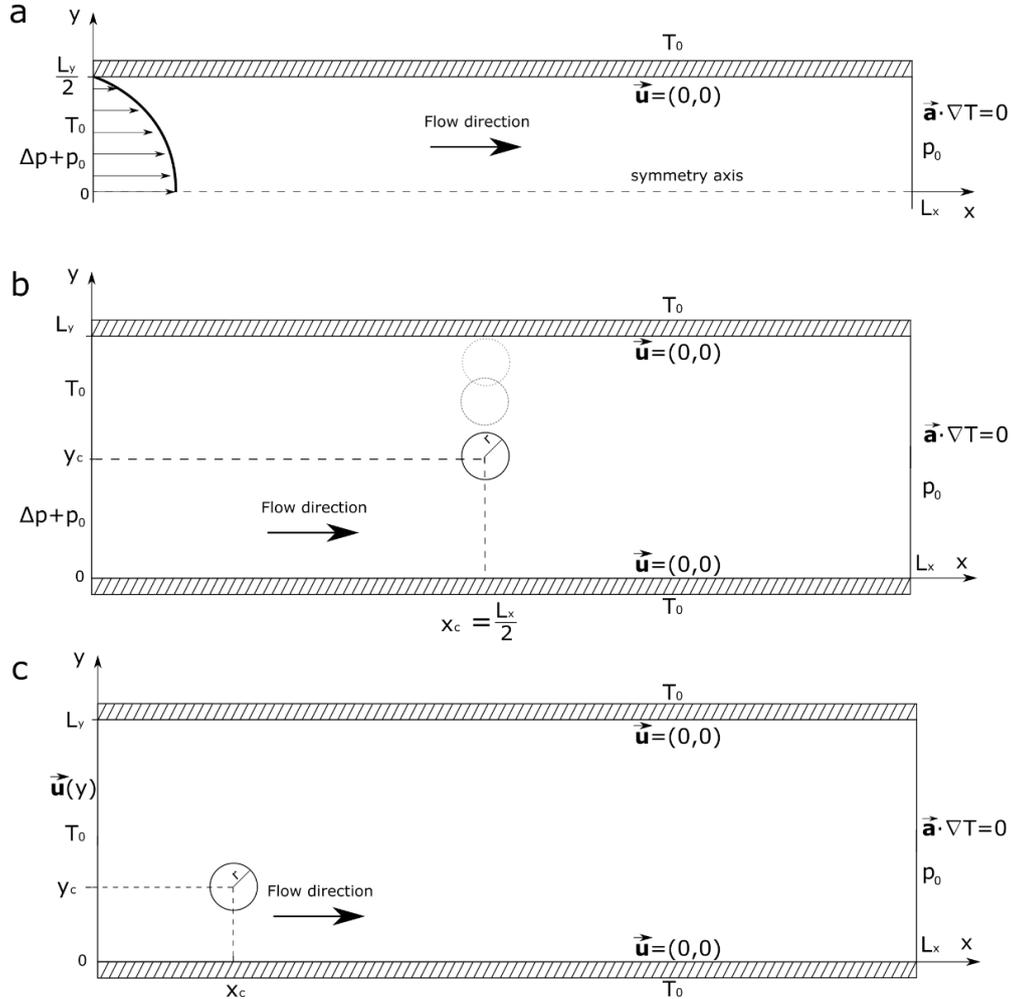

FIG. 1. Schematic representation of the geometry of the simulated 2-dimensional systems along with the imposed boundary conditions. (a) Poiseuille flow with parabolic velocity profile. (b) flow past a cylindrical obstacle fixed at different positions in the direction perpendicular to the flow. (c) Unsteady, oscillatory flow past a cylindrical object with a fixed position at $x_c = 0.2$ m, $y_c = 0.2$ m. The dimensions of the channel were $L_x = 2.2$ m and $L_y = 0.41$ m. In all the cases the system was translationally invariant in the z-direction with a fixed depth equal to $L_z = 1$ m.

In case (i), the system was meshed with 10000 (100×100) rectangular elements. In the *x* direction, equal-sized elements were distributed uniformly. In the *y* direction, they were distributed in an arithmetic sequence. The element size ratio of the largest element in the sequence to the smallest one was set to 4. In cases (ii) and (iii), we applied triangular, physics-controlled mesh due to the presence of circular domains in the geometry. We used the default 'extra fine' option with a maximal element size equal to (ii) $3.35 \cdot 10^{-4}$ m and (iii) $2.75 \cdot 10^{-4}$ m. For details on meshing in COMSOL see Chapter 8 from ref. [29]. The relative tolerance of the solver was set to $10^{-5}$ in all studies. Validation of the model (i) against a simplified analytical solution was performed in our previous paper [16]. Based on this study, we estimated the absolute error of *U* measurements to $2.8 \cdot 10^{-9}$ J. This estimation was done based on the mesh refinement study for $\Delta p = 0.163$ Pa as described in ref. [16].

### III. RESULTS

In this work, we explore the transfer of a system's internal energy to the surrounding through mechanisms such as heat, matter, and work. Our recent series of publications have demonstrated, through analytical examples, that these quantities can be directly computed from boundary conditions, leading to averaged quantities via a mapping procedure described previously [19–21]. Here, we consider more complicated scenarios where an indirect, numerical approach is needed. In such cases, we first calculate pressure and temperature profiles ($\boldsymbol{p(x,y)}, \boldsymbol{T(x,y)}$). From the equations of state for ideal gas and local equilibrium assumption we get the following density profile:

$$\rho(x,y) = \frac{Mp(x,y)}{RT(x,y)}. \tag{8}$$

Based on these profiles from each simulation, we obtain the total internal energy of the gas and the number of moles using the following integrals over the simulation domain *A*:

$$U = \int \frac{3R\rho(x,y)T(x,y)L_z}{2M} dA, \tag{9}$$

$$n = \int \frac{\rho(x,y)L_z}{M} dA. \tag{10}$$

Using these quantities, we can calculate the effective NET parameters such as temperature, pressure, chemical potential and entropy (indexed with a star):

$$T^* = \frac{2U}{3nR}, \quad p^* = \frac{2U}{3V}, \quad \mu^* = \frac{5}{2}RT^* - \frac{5}{2}RT^*\ln\left(\frac{T^*}{T_0^*}\right) + RT^*\ln\left(\frac{p^*}{p_0^*}\right), \quad S^* = \frac{3}{2}nR\ln\left(\frac{Un_0}{nU_0}\right) + nR\ln\left(\frac{Vn_0}{nV_0}\right).$$

In all the following studies, we changed one control parameter while keeping others constant. We considered the first point in a series of simulations as the reference point from which we calculated the integrated NET differentials: $\Delta U$, $\int \mu^* dn$, $\int T^* dS^*$ and $\int p^* dV$. For example, in simulations with $T_0$ varying from 600 K to 610 K, the initial values of calculated quantities were $U_0 = U(600K)$, $n_0 = n(600K)$ etc.

## A. Planar Poiseuille flow

We started the analysis by performing 75 simulations of the flow, wherein we systematically varied the pressure difference $\Delta p$ across a range from 0.002 to 0.15 Pa. The points were uniformly distributed, with intervals of 0.002 Pa. Throughout these simulations, we maintained the remaining control parameters at constant values, specifically $T_0 = 600$ K and $L_y = 0.05$ m. In Figure 2a-c, we present three integrated NET differentials as defined in the preceding section. The absolute change in these quantities is consistently on the order of $10^{-3}$ J. Notably, in Figure 2d, we demonstrate that their sum is effectively 0, accounting for numerical errors within the computational framework.

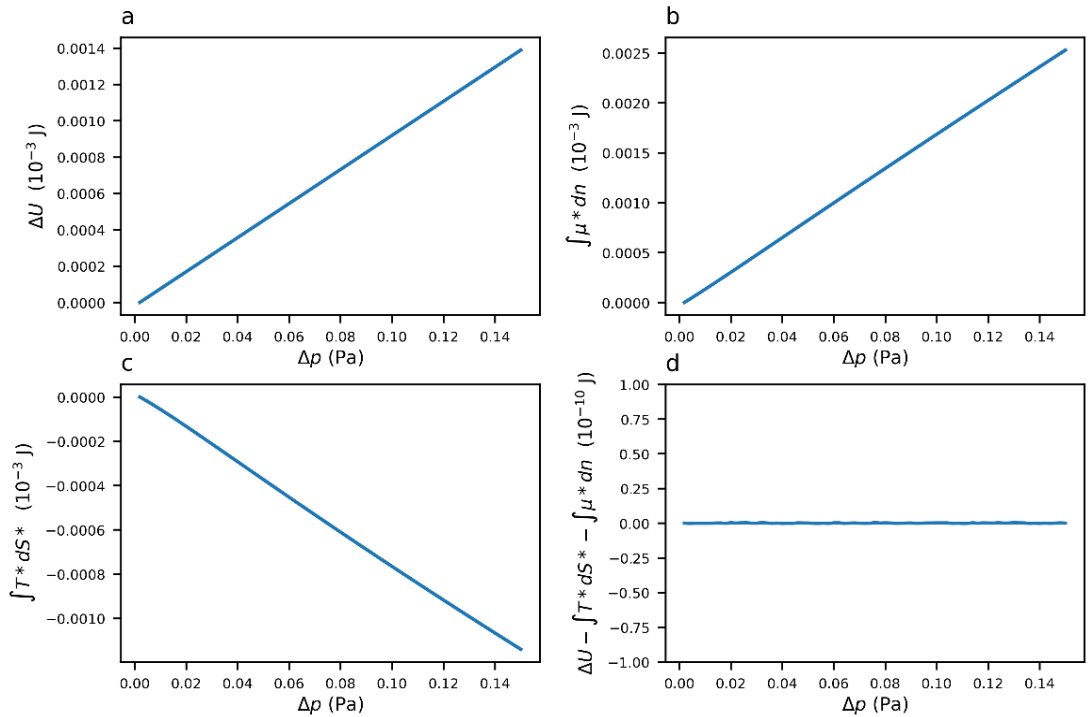

FIG. 2. (a) Variation of integrated NET differentials a) $\Delta U$, b) $\int \mu^* dn$, c) $\int T^* dS^*$ and d) $\Delta U - \int \mu^* dn - \int T^* dS^*$ as a function of $\Delta p$ in the range of 0 to 0.15 Pa. The values of the other control parameters were $T_0 = 600$ K and $L_y = 0.05$ m. Please note that the scale in Fig. 2d is in the units of $10^{-10}$, thus the difference $\Delta U - \int \mu^* dn - \int T^* dS^* = 0$ is within numerical error. The maximal absolute errors are equal to $2.8 \cdot 10^{-9}$ J.

Subsequently, we conducted a comparable set of simulations, extending our investigation to $\Delta p$ values ranging from 1 to 70 Pa. This range encompassed 70 equidistant data points, systematically spaced at 1 Pa intervals. Note that for the pressure difference shown in Figure 2, Reynolds numbers remained constrained ($Re \leq 382$) corresponding to the laminar flow regime. For the increase pressure range shown in Figure 3 the Reynolds number reaches 147,000 at $\Delta p = 70$ Pa, resulting in unsteady flow in actual conditions. Here, we do not consider roughness of the wall or any channel imperfections. Thus, solving stationary equations, even at such high $Re$ values, led to laminar flow solutions. At the increased $\Delta p$ the integrated differentials associated with mass and heat exchange, shown in Figure 3b and 3c, respectively, exhibited nonlinearity. Their respective absolute changes are 0.8 J and 0.11 J, while the sum of all the terms remained 0, as indicated in Figure 3d.

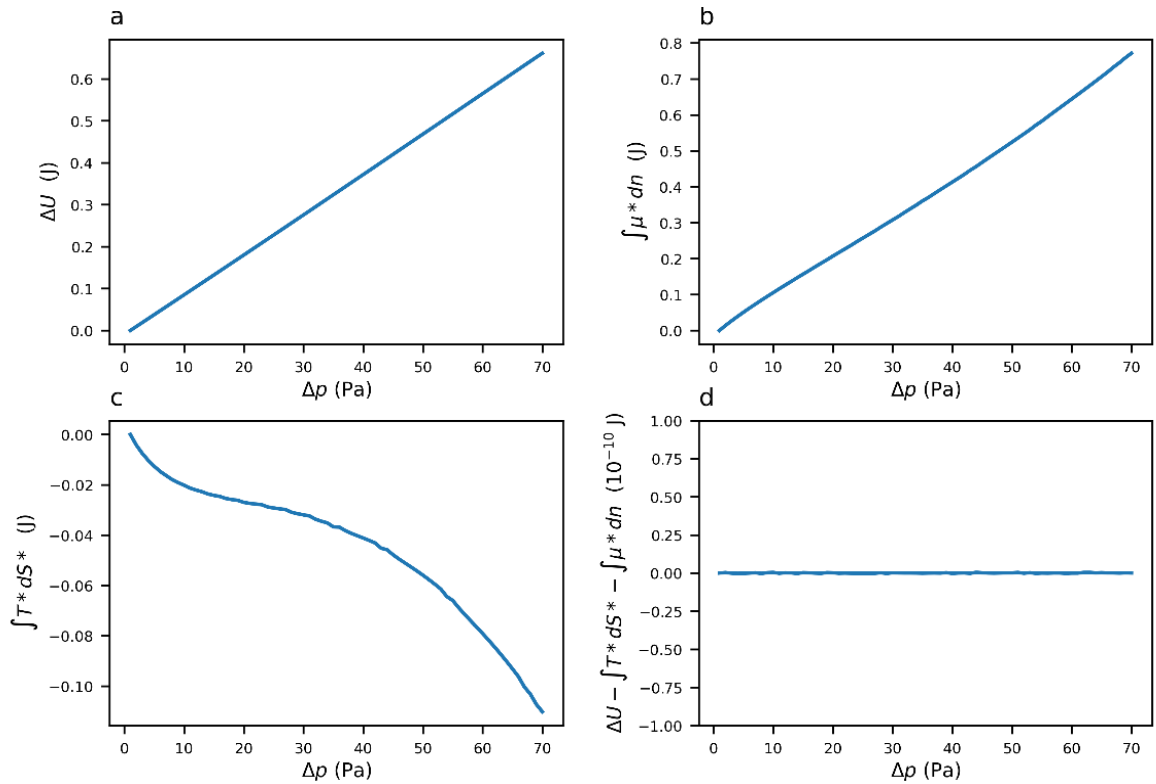

FIG. 3. Variation of integrated NET differentials a) $\Delta U$, b) $\int \mu^* dn$, c) $\int T^* dS^*$ and d) $\Delta U - \int \mu^* dn - \int T^* dS^*$ as a function of $\Delta p$ in the range of 0 to 70 Pa. The values of the other control parameters were $T_0 = 600$ K and $L_y = 0.05$ m. Please note that the scale in Fig. 3d is in the units of $10^{-10}$, thus the difference $\Delta U - \int \mu^* dn - \int T^* dS^* = 0$ is within numerical error. The maximal absolute errors are equal to $2.8 \cdot 10^{-9}$ J.

In the next analysed process, we modulated $T_0$ within the range of 600.01 to 610 K while maintaining other parameters constant at $\Delta p = 0.02$ Pa and $L_y = 0.05$ m. A comprehensive series of 1000 simulations was conducted, systematically varying $T_0$ by increments of 0.01 K. Changes in integrated NET differentials are elucidated in Figure 4. It can be seen that during this process, not only does the sum of all terms equal zero, but the internal energy $U$ remains constant with respect to variations in the boundary temperature. This phenomenon is explicable through the equation of state, wherein a corresponding decrease in density counteracts the temperature increase. Consequently, the overall internal energy of the system undergoes no net change. Note that the entropic and mass transfer contributions are significant compared to the previously considered pressure-controlled process (see Figure 2 and Figure 3).

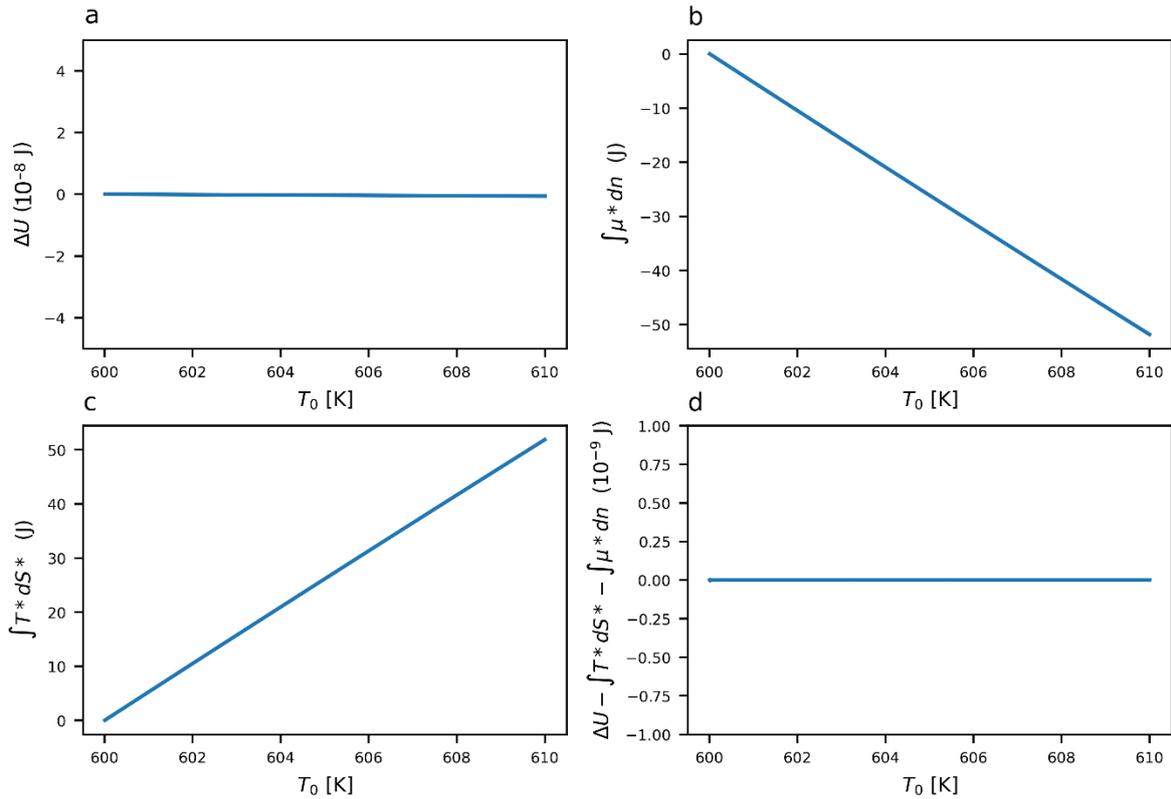

FIG. 4. Variation of integrated NET differentials a) $\Delta U$, b) $\int \mu^* dn$, c) $\int T^* dS^*$ and d) $\Delta U - \int \mu^* dn - \int T^* dS^*$ in function of $T_0$. The value of the other control parameters were $\Delta p = 0.02$ Pa and $L_y = 0.05$ m. Please note that the scale in Fig. 4d is in the units of $10^{-6}$, thus the difference $\Delta U - \int \mu^* dn - \int T^* dS^* = 0$ is within numerical error. The maximal absolute errors are equal to $2.8 \cdot 10^{-9}$ J.

When the system undergoes a volume change, an additional term associated with work performed on or by the system emerges in the context of the first law of thermodynamics. To validate the persistence of the formulated NET first law in such circumstances, we conducted a series of 46 simulations wherein the volume of the channel was systematically altered within the range of 0.0025 to 0.025 m³. Note that only $L_y$ dimension of the channel was changed (ranging from 0.01 to 0.1 m) while $L_x$ and $L_z$ remained constant. In Figure 5, we demonstrate the change of the integrated NET differentials and their sum. Note that in this case, there is an additional term $\int p^* dV$ (Figure 5d) corresponding to the work associated with the flow. The magnitudes of the absolute changes in $\Delta U$, $\int \mu^* dn$ (also, the work term $\int p^* dV$) are substantial when compared to the preceding processes. The changes in the heat transfer term are eight orders of magnitude smaller. Interestingly, this term is non-monotonic and has a minimum at $V = 0.0205$ m³. Crucially, the total sum of these integrated differentials is 0, as illustrated in Figure 5e, accounting only for numerical error.

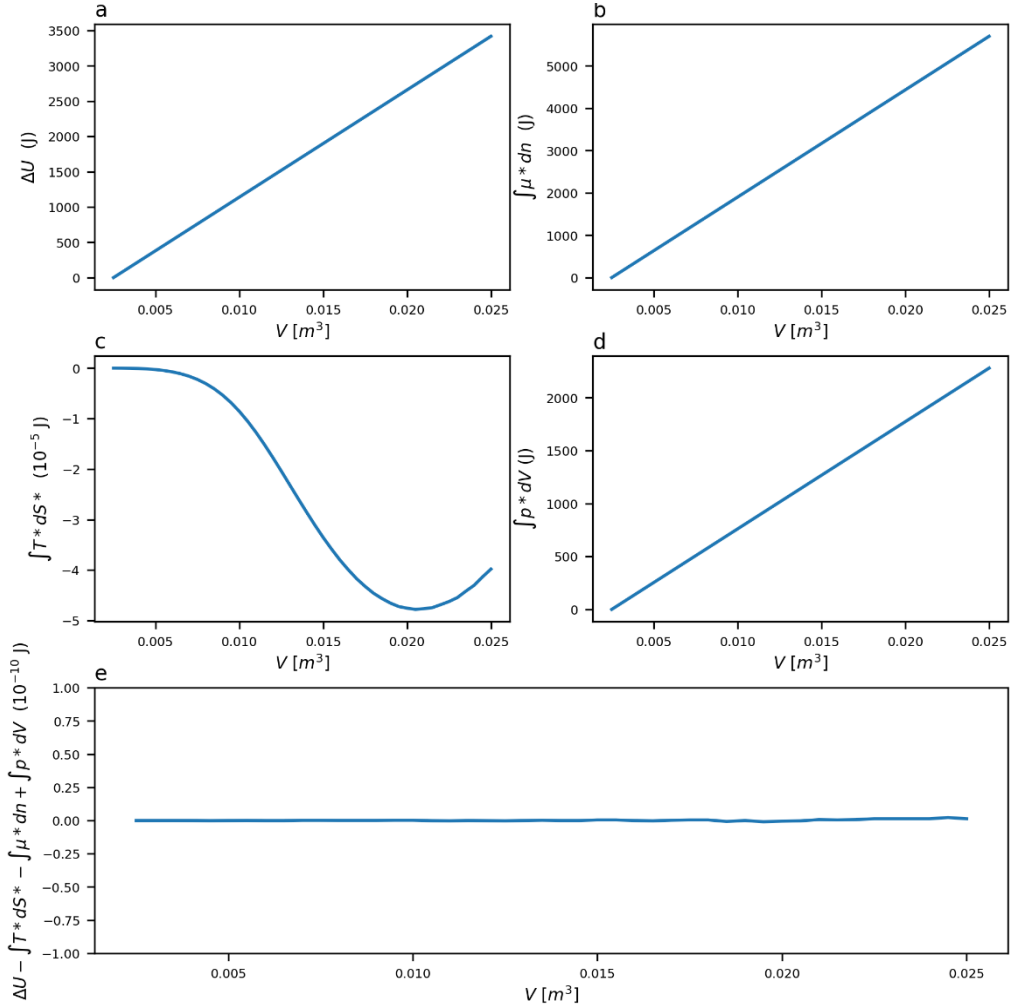

FIG. 5. Variation of integrated NET differentials a) $\Delta U$, b) $\int \mu^* dn$, c) $\int T^* dS^*$, d) $\int p^* dV$ and e) $\Delta U - \int \mu^* dn - \int T^* dS^* + \int p^* dV$ in function of $V = \frac{L_x L_y L_z}{2}$. To change $V$ we varied $L_y$ from 0.01 to 0.1 m. The value of the other control parameters were $\Delta p = 0.02$ Pa, $T_0 = 600$ K. Please note that the scale in Fig. 5e is in the units of $10^{-10}$, thus the difference $\Delta U - \int \mu^* dn - \int T^* dS^* = 0$ is within numerical error. The maximal absolute errors are equal to $2.8 \cdot 10^{-9}$ J.

## B. Cylinder moved in y direction

In this study, we conducted three sets of 40 simulations. In each simulation series, a cylinder was displaced in the y-direction from its initial position at the centre of the channel (at $x_c = \frac{L_x}{2}$, $y_c = $

$\frac{L_y}{2}$) towards the wall, with its final coordinates being $x_c = \frac{L_x}{2}$, $y_c = \frac{9}{80}L_y$. The varying factor among the series was $\Delta p$, with values set at 0.01 Pa, 0.02 Pa and 0.03 Pa; the other control parameters ($T_0$, $L_x$, $L_y$, $L_z$, $p_0$) remained constant across all simulations. Figure 6 illustrates the impact of the cylinder positioning on pressure and velocity profiles in the flow. As anticipated, the obstacle at the centre, exposed to higher fluid velocities, induces more significant modifications in the flow field. Specifically, the central obstacle blocks the flow more efficiently, which is visible as the abrupt pressure drop compared to the near-wall positioning of the obstacle. Consequently, the average velocity for the first case is significantly smaller than the second one.

At every position of the cylinder, we calculated the integrated NET differentials of the flow as previously, and we demonstrated their changes in Figure 7. We found that these are orders of magnitude smaller than the processes performed for pure Poiseuille flow at similar $\Delta p$ (cf. Figure 2). This proves that cylinder shifting is a smoother process regarding modifications of the integrated NET differentials. On the other hand, the differences can be pronounced when a higher pressure is applied. Interestingly, at $\Delta p = 0.01$ Pa, shifting the cylinder from the centre of the channel to $y_c = 0.04$ m leads to positive values of energy exchanged in the form of heat $\int T^* dS^* > 0$. Nevertheless, irrespectively of $\Delta p$ the sum of all integrated differentials, presented in Figure 7d is 0 within the numerical error, indicating that the postulated first non-equilibrium thermodynamic rule is valid in a process with all control parameters fixed and the flow is altered only due to geometric constraint.

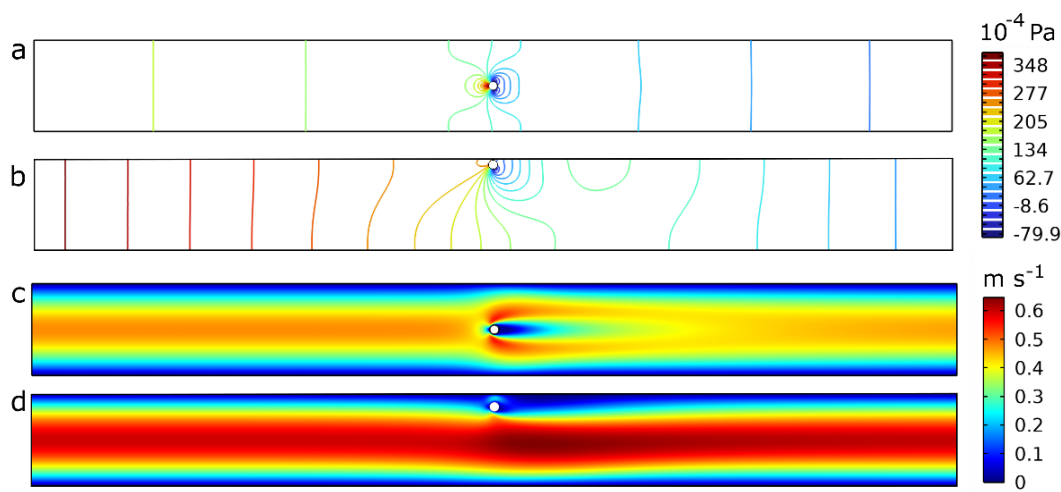

FIG. 6. (a, b) Pressure and (c, d) velocity profiles of the compressible flow of ideal gas past a cylinder placed at (a,c) the centre of the channel and (b,d) close to its wall ($y_c = 0.043$ m).

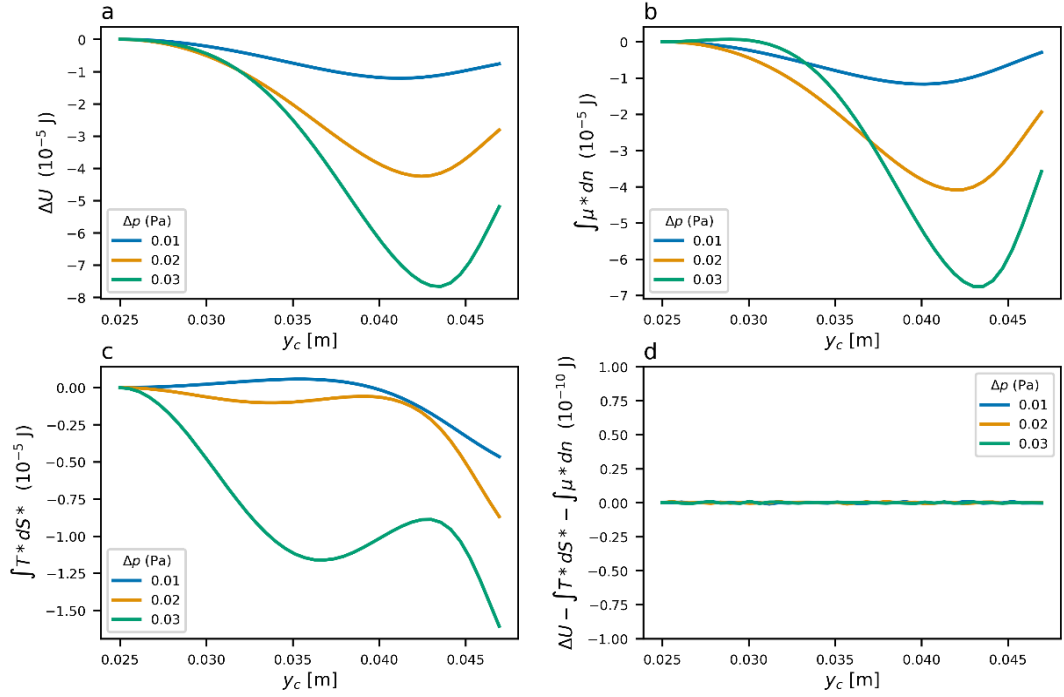

FIG. 7. Variation of integrated thermodynamic differentials, namely a) $\Delta U$, b) $\int \mu^* dn$, c) $\int T^* dS^*$ and d) $\Delta U - \int \mu^* dn - \int T^* dS^*$ in function of y position of the cylinder $y_c$ at three values of $\Delta p$. The value of the other control parameters were $T_0$ = 273.15 K and $L_y$ = 0.05 m. Please note that the scale in Figure 7d is in the units of $10^{-10}$, thus the difference $\Delta U - \int \mu^* dn - \int T^* dS^* = 0$ is within numerical error. The maximal absolute errors are equal to $2.8 \cdot 10^{-9}$ J.

## C. Oscillatory flow

Finally, we analysed an unsteady flow with von Karman vertices. In this study, we observed the time evolution of a system defined in a benchmark study [28]. In contrast to the previous cases, we fixed all control parameters and observed time variations of the integrated NET differentials. We first traced the flow's kinetic energy to identify the onset of stable oscillations. In the beginning, the system was developing, steadily increasing its kinetic energy, as shown in Figure 8a. At $t$ = 2 s, the flow became oscillatory with von Karman vortices visible in the velocity profile (see Figure 8b). For analysis of the integrated thermodynamic potentials, we selected the pure oscillatory stage from $t$ = 2.48 to 2.7 s with kinetic energy changes visible in the inset. The corresponding NET integrals are given in Figure 9. The values of $\Delta U$, $\int \mu^* dn$ and $\int T^* dS^*$ oscillate with the same period as kinetic energy, yet their total sum equals 0 within numerical error, proving that the first rule of non-equilibrium thermodynamics also applies to non-stationary conditions.

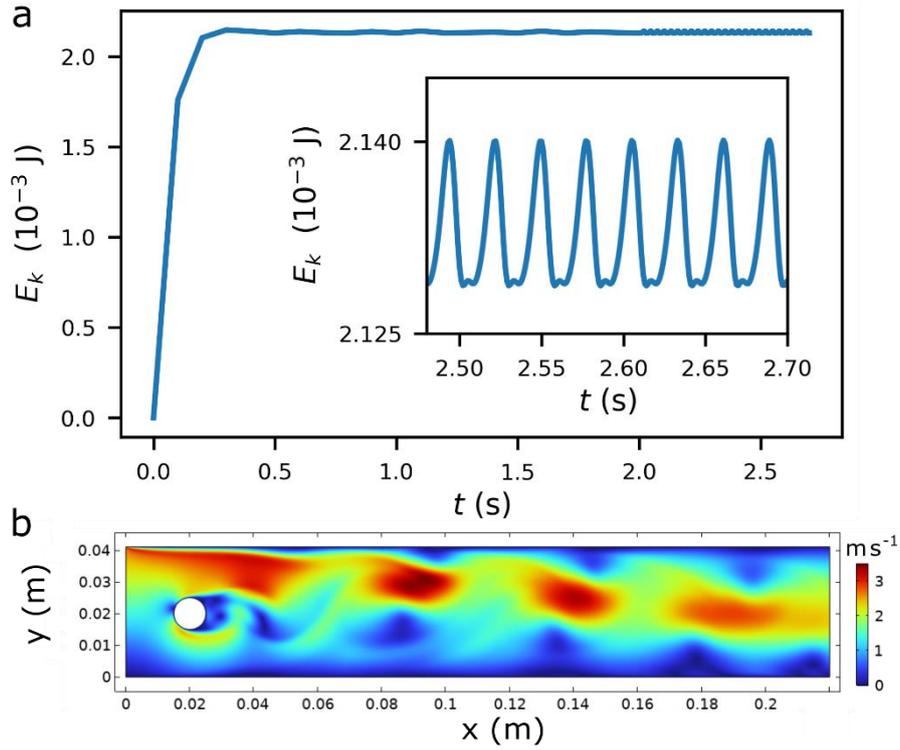

FIG. 8. (a) Variation of kinetic energy in time. Stable oscillations from 2.48 to 2.7 s were selected for the the analysis of the integrated NET differentials. (b) Snapshot of the velocity field at t = 2.6 s.

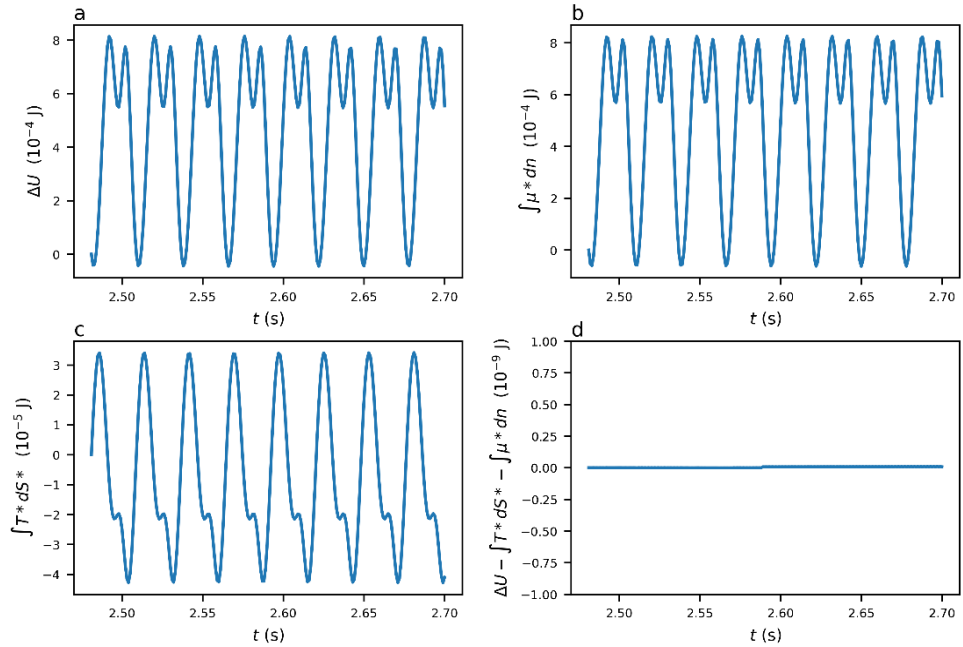

FIG. 9. Variation of integrated NET differentials a) $\Delta U$, b) $\int \mu^* dn$, c) $\int T^* dS^*$ and d) $\Delta U - \int \mu^* dn - \int T^* dS^*$ in function of time. The value of the other control parameters were $T_0 = 600$ K and $p_0 = 1$ atm. Please note that the scale in Figure 9d is in the units of $10^{-9}$, thus the difference $\Delta U - \int \mu^* dn - \int T^* dS^* = 0$ within numerical error. The maximal absolute errors are equal to $2.8 \cdot 10^{-9}$ J.

## IV. CONCLUSIONS

We have analysed the Poiseuille flow of ideal gas between two parallel plates kept at a fixed temperature. Additionally, we placed a cylinder between the plates and observed steady and unsteady flow in this system. We changed the temperature of the plates, the distance between them, the pressure difference between inlet and outlet and the location of the cylinder. We followed the changes of the internal energy in these processes and confirmed that $U(S^*, V, N)$ (see Figures 2-5 and 7) depends solely on the non-equilibrium entropy, volume and number of particles in the system. We increased the flow rate past the cylinder, and the flow became unsteady. The temperature, density, pressure, internal energy and kinetic energy oscillated as a function of time, t. We investigated $U$ as a function of time, t, only via the parameters of state $U(S^*(t), V, N(t))$. Thus, we showed that such a form of internal energy is robust, irrespective of the total number of external parameters affecting the steady or unsteady flow. Previously, we have obtained the same form of $U$ for the ideal gas subjected to heat flow only [23]. Here, we added a flow of matter past the system and examined the same functional form of $U$. We conclude that, irrespective of how the system exchanges energy with the surroundings, the energy of one-component ideal gas is given by three global parameters of state.

## ACKNOWLEDGMENTS

This research was founded by NCN within Sonata grant 2019/35/D/ST5/03613.